\documentclass[aps,letterpaper,twocolumn,noshowpacs,preprintnumbers,amsmath,amssymb,showkeys]{revtex4}

\usepackage{epsfig}
\usepackage{graphicx}
\usepackage{dcolumn}
\usepackage{bm}
\usepackage{bbm}

\def \ud {\dot{u}}
\def \vd {\dot{v}}
\def \zd {\dot{z}}

\def \Ud {\dot{U}}
\def \Zd {\dot{Z}}
\def \Cd {\dot{C}}
\def \Xd {\dot{X}}
\def \Pd {\dot{P}}

\def \ab {\bar{a}}
\def \ub {\bar{u}}
\def \vb {\bar{v}}
\def \zb {\bar{z}}

\begin{document}


\title{Fast Correlation Greeks by Adjoint Algorithmic Differentiation}

\author{Luca Capriotti $^1$} \email{luca.capriotti@credit-suisse.com.} 

\author{Mike Giles $^2$ }\email{mike.giles@maths.ox.ac.uk}

\affiliation{%
$^1$ Quantitative Strategies, Investment Banking Division, Credit Suisse Group,\\
Eleven Madison Avenue, New York City, NY 10010-3086, United States of America \\
$^2$ Oxford-Man Institute of Quantitative Finance and Oxford University Mathematical Institute,\\
24-29 St. Giles, Oxford, OX1 3LB, United Kingdom 
}%

\date{\today}

\begin{abstract}
We show how Adjoint Algorithmic Differentiation (AAD) allows an extremely efficient 
calculation of correlation Risk of option prices computed with Monte Carlo 
simulations. 
A key point in the construction is the use of binning to simultaneously achieve
computational efficiency and accurate confidence intervals.
We illustrate the method for a copula-based Monte Carlo computation
of claims written on a basket of underlying assets, and we test it numerically 
for Portfolio Default Options.  For any number of underlying assets or names in a portfolio, 
the sensitivities of the option price with respect to all the
pairwise correlations is obtained at a computational cost which is at most 4 times
the cost of calculating the option value itself.  For typical 
applications, this results in computational savings of several order of magnitudes
with respect to standard methods. 
\end{abstract}

\keywords{Algorithmic Differentiation, Monte Carlo Simulations, Derivatives Pricing, Credit Derivatives}

\maketitle

One of the consequences of the current crisis of the Financial Markets
is a renewed emphasis on rigorous Risk management practices. 
In order to quantify the financial exposure of financial firms, 
and to ensure an efficient capital allocation, and more effective hedging practices, 
regulators and senior management alike are insisting more and more on a 
thorough monitoring of Risk. Among all businesses, those dealing with complex, 
over the counter derivative securities are the ones receiving the most attention.

A thorough calculation of the Risk exposure of portfolios of structured derivatives
comes with a high operational cost because of the large amount of computer power required. 
Indeed, highly time consuming Monte Carlo (MC) simulations are very often the only tool 
available for pricing and hedging complex securities. 
Calculating the Greeks, or price sensitivities, by `Bumping' i.e., by 
perturbing in turn the underlying model parameters, repeating the simulation and forming finite difference
approximations  results in a computational burden increasing linearly with the number of sensitivities
computed. This easily becomes very significant when the models employed depend on a large number
of parameters, as it is typically the case.

A particularly challenging task is the calculation of correlation Risk, i.e., the calculation 
of the sensitivites of a security with respect to some measure of the correlations 
among the random factors it depends on. Indeed, 
calculating Risk with respect to all the independent pairwise correlations by Bumping requires repeating the 
MC simulation a large number of times, e.g. increasing quadratically with the number of random
factors, and it is often unfeasible because of its high computational cost. 

Several alternative methods for the calculation of price sensitivities have been proposed in the literature (for a review see e.g., \cite{GlassMCbook}). 
Among these, the Pathwise Derivative method \cite{BrodGlass} provides unbiased estimates at a computational cost  that may be smaller than the one of Bumping. However,
in many problems the standard Pathwise Derivative method provides limited computational gains, especially when the contract priced has a complex payout \cite{aad1}.
A much more efficient implementation of the Pathwise Derivative method was proposed by Giles and Glasserman in Ref.~\cite{GilesGlasserman} in the context of the Libor Market Model for
European payouts, and recently generalized to Bermudan options by Leclerc and co-workers \cite{Leclerc}. These formulations express the calculation of the Pathwise Derivative 
estimator in terms of linear algebra operations, and utilize adjoint methods to reduce the computational complexity by rearranging appropriately the 
order of the calculations. 

Adjoint implementations can be seen as instances of a programming technique known as Adjoint 
Algorithmic Differentiation (AAD) \cite{griewank, GilesProc}. 
In particular, as also discussed in a forthcoming paper \cite{aad2}, AAD can be used as a {\em design paradigm} to implement the Pathwise Derivative method, or the 
calculation of the sensitivities of any numerical algorithm, in full generality.  
In this paper we illustrate these ideas by discussing a specific application:  the calculation of correlation Risk.
We will begin by introducing the main ideas underlying Algorithmic Differentiation (AD), and the results on the computational 
efficiency of its two basic approaches: the Forward and Adjoint modes.

\section*{Forward and Adjoint Algorithmic Differentiation}

Both the Forward and Adjoint mode of AD aim at calculating the derivatives of 
a computer implemented function. They differ by the direction of 
propagation of the chain rule through the composition of instructions representing the function.
To illustrate this point, suppose we begin with a single input $a$, and produce a single output 
$z$ after proceeding through a sequence of steps:
\[
a\ \rightarrow\ \ldots\ \rightarrow\ u\ \rightarrow\ v\ \rightarrow\ \ldots\ \rightarrow\ z.
\]

The Forward (or Tangent) mode of AD (FAD) defines $\ud$ to be the sensitivity 
of $u$ to changes in $a$, i.e.,
\[
\ud \equiv \frac{\partial u}{\partial a}~.
\]
If the intermediate variables $u$ and $v$ are vectors, $\vd$ is
calculated by differentiating the dependence of $v$ on $u$ so that
\[
\vd_i = \sum_j \frac{\partial v_i}{\partial u_j}\ \ud_j.
\]
Applying this to each step in the calculation, working from left to right, 
we end up computing $\zd$, the sensitivity of the output to changes in the input.
Note that if we have more than one input, then we need to calculate the sensitivity 
to each one in turn, and so the cost is linear in the number of input variables.

Instead, the Adjoint (or Backward) mode of AD (AAD) works from right to left.  Using the 
standard AD notation, $\ub$ is defined to be the sensitivity of the output $z$ to 
changes in the intermediate variable $u$, i.e.
\[
\ub_i \equiv \frac{\partial z}{\partial u_i}.
\]
Using the chain rule we get,
\[
\frac{\partial z}{\partial u_i} = \sum_j \frac{\partial z}{\partial v_j}\ 
\frac{\partial v_j}{\partial u_i},
\]
which corresponds to the adjoint mode equation
\[
\ub_i = \sum_j \frac{\partial v_j}{\partial u_i}\ \vb_j.
\]
Starting from $\zb=1$, we can apply this to each step in the calculation, working from 
right to left, until we obtain $\ab$, the sensitivity of the output to each of the input 
variables.

In the Adjoint mode, the cost does not increase with the number of inputs, but if
there is more than one output then the sensitivities for each output have to considered
one at a time and so the cost is linear in the number of outputs.
Furthermore, because the partial derivatives depend on the values of the intermediate 
variables, one first has to compute the original calculation storing the values of
all of the intermediate variables such as $u$ and $v$, before performing the Adjoint mode 
sensitivity calculation.

In the above description, each step can be a distinct high-level function, 
or specific mathematical operations, 
or even an individual instruction in a computer code.
This last viewpoint is the one taken by computer scientists who have developed tools 
which take as an input a computer code to perform some high-level function, 
\[
V = \texttt{FUNCTION}(U)
\]
and produce new routines which will either perform the standard sensitivity analysis
\[
\dot V = \texttt{FUNCTION}\_{\texttt D}(U, \dot U)
\]
with suffix $\_{\texttt D}$ for ``dot'', or its adjoint counterpart
\[
\bar U = \texttt{FUNCTION}\_{\texttt B}(U, \bar V)
\]
with suffix $\_{\texttt B}$ for ``bar'' 
\footnote{To learn more about Automatic Differentiation tools
see e.g., \texttt{www.autodiff.org}. }.  

One particularly important theoretical result is that 
the number of arithmetic operations in the adjoint routine $\texttt{FUNCTION}\_{\texttt B}$ is 
at most a factor 4 greater than in $\texttt{FUNCTION}$ \cite{griewank}. As a result, it is 
possible to show that the execution time of $\texttt{FUNCTION}\_{\texttt B}$  is bounded by approximatively 4 times the cost of 
execution of the original function $\texttt{FUNCTION}$.
Thus, one can obtain the sensitivity of a single output to
an unlimited number of inputs for little more work than the original computation.

While the application of such {\em automatic} AD tools to large inhomogeneous
pricing softwares is challenging, the principles of AD can
be used as a programming paradigm that can be used to design the Forward or Adjoint of
any algorithm (possibly using automatic AD tools for the implementation of smaller, simpler 
components).  This is especially useful for the most common situations where pricing codes use a 
variety of libraries written in different languages, possibly linked dynamically.
These ideas will be discussed at length in Ref.~\cite{aad2}.
 
\section*{AAD and the Pathwise Derivative method for Correlation Risk}

In this paper, we consider options pricing problems that can be expressed as an expectation value
of the form
\begin{equation}\label{option}
V = \mathbb{E}_\mathbb{Q}\Big[P(X)\Big]~,
\end{equation}
where $X = (X_1,\ldots,X_N)^t$ represents the state vector of $N$ market factors
(e.g., stock prices, interest rates, foreign exchange pairs, default times etc.), 
$P(X)$ 
is the (possibly discounted) payout function of a security contingent on their future realization,
and $\mathbb{Q} = \mathbb{Q}(X) $ represents 
a risk neutral probability distribution \cite{HarrKreps} according to 
which the components of $X$ are distributed.
Although the proposed method easily generalizes to other kinds of joint distributions, here
we consider a $N$-dimensional Gaussian copula as a model for the co-dependence between 
the components of the state vector, namely a joint cumulative density function of the form
\begin{equation} \label{gausscopula}
\mathbb{Q}(X) = \Phi_N(\Phi^{-1}(M_1(X_1)),\ldots, \Phi^{-1}(M_N(X_N));\rho)
\end{equation}
where $\Phi_N(Z_1,\ldots,Z_N;\rho)$ is a $N$-dimensional multivariate Gaussian distribution
with zero mean, and a $N\times N$ positive semidefinite correlation 
matrix $\rho$; $\Phi^{-1}$ is the inverse of the standard
normal cumulative distribution, and $M_i(X_i)$, $i=1,\ldots,N$, are the 
Marginal distributions of the underlying factors, typically implied from the market prices of
liquid securities.  

The expectation value in (\ref{option}) can be estimated by means of
MC by sampling a number $N_{\rm MC}$ of random
replicas of the underlying state vector ${X}[1],\ldots,{X}[N_{\rm MC}]$, 
according to the distribution $\mathbb{Q}({ X})$, and evaluating the payout $P({X})$
for each of them. This leads to the central limit theorem \cite{CLT} estimate of the option value $V$
as
\begin{equation}\label{mcexp}
V \simeq \frac{1}{N_{\rm MC}} \sum_{i_{\rm MC} = 1}^{N_{\rm MC}} P\left({X}[i_{\rm MC}]\right)
\end{equation}
with standard error $\Sigma / \sqrt{N_{\rm MC}}$,
where $\Sigma^2 = E_{\mathbb Q}[P\left({X}\right)^2]-E_{\mathbb Q}[P\left(X\right)]^2$ is the 
variance of the sampled payout.

In the Gaussian model above the dependence between the underlying factors is determined by the 
correlation of a set of jointly normal random variables $Z = (Z_1,\ldots,Z_N)^t$ distributed according 
to $\Phi_N(Z_1,\ldots,Z_N;\rho)$. Each $Z_i$ is distributed according to a standard normal distribution 
so that $\Phi(Z_i)$ is a uniform random variable in $(0,1)$ and $X_i = M_i^{-1}(\Phi(Z_i))$ is 
distributed according to $M_i$.  The sampling of the $N$ jointly normal random variables $(Z_1,\ldots,Z_N)$
is efficiently implemented by means of a Cholesky factorization of the correlation matrix. The
Cholesky factorization produces a lower triangular $N\times N$ matrix $C$ such that $\rho=CC^T$ so 
that one can write $Z = C \tilde Z$ where $\tilde Z = (\tilde Z_1, \ldots, \tilde Z_N)^t$ is a 
$N$ dimensional vector of independent standard normal random variables.  These observations naturally translate into the 
standard algorithm to generate MC samples of $X$ according to (\ref{gausscopula}), namely
\begin{itemize} 
\item[Step 0] Generate a sample of $N$ independent standard normal variates, $\tilde Z = (\tilde Z_1, \ldots, \tilde Z_N)^t$.
\item[Step 1] Correlate the components of $\tilde Z$ by performing the matrix vector product $Z = C \tilde Z$. 
\item[Step 2] Set $U_i = \Phi(Z_i)$, $i=1,\ldots, N$.
\item[Step 3] Set $X_i = M^{-1}(U_i)$, $i=1,\ldots, N$.
\item[Step 4] Compute the payout estimator $P(X_1,\ldots,X_N)$.
\end{itemize}

Correlation Risk can be obtained in an highly efficient way by implementing the so-called Pathwise Derivative 
method \cite{BrodGlass} according to the principles of AAD \cite{aad1,aad2}.
It is convenient to first express the expectation value as being over $\mathbb{P}(\tilde Z)$, 
the distribution of independent $\tilde Z$ used in the MC simulation, so that
\begin{equation}
V = 
\mathbb{E_Q}\Big[P\left({X}\right)\Big] =
\mathbb{E_P}\Big[P\left({X(\tilde Z)}\right)\Big].
\label{option2}
\end{equation}
The point of this subtle change is that $\mathbb{P}(\tilde Z)$ does not depend on the correlation matrix 
$\rho$, whereas $\mathbb{Q}(X)$ does.

The Pathwise Derivative method allows the calculation of the sensitivities of the option price $V$ (\ref{option2}) 
with respect to a set of  $N_\theta$ parameter $\theta = (\theta_1,\ldots, \theta_{N_\theta})$, say 
\begin{equation}\label{sens}
 \frac{\partial V(\theta)}{\partial\theta_k} =
 \frac{\partial}{\partial \theta_k} \mathbb{E_P}\Big[P\left({X}\right)\Big]~,
\end{equation}
by defining appropriate estimators, say $\bar \theta_k(X[i_{MC}])$, that can be sampled simultaneously 
in a single MC simulation. This can be achieved by observing that whenever the payout function is regular enough (e.g., Lipschitz-continuous, see
Ref.~\cite{GlassMCbook}), and the distribution $\mathbb{P}(\tilde Z)$ does not depend on $\theta$, 
one can rewrite Eq.~(\ref{sens}) by taking the derivative inside the expectation value, as
\begin{equation}\label{sens2}
 \frac{\partial V(\theta)}{\partial\theta_k} = \mathbb{E_P}\Big[\frac{\partial P \left({X}\right) }{\partial \theta_k} \Big]~.
\end{equation}

The calculation of Eq.~(\ref{sens2}) can be performed by applying the chain rule, and 
computing the average value of the so-called Pathwise Derivative estimator
\begin{equation}\label{pwd}
\frac{\partial P(X)}{\partial \theta_k} = 
\sum_{i=1}^{N} \frac{\partial P(X)}{\partial X_i}
\times\frac{\partial X_i}{\partial \theta_k}  ~.
\end{equation}

The standard pathwise implementation corresponds to a Forward mode sensitivity analysis.
Applied to steps 1-4 (since the normal variates $\tilde Z$ do not depend on any
input parameters), this gives for each sensitivity:

\begin{itemize}
\item[Step 1f] Calculate $\Zd = \Cd\, \tilde Z$ where $\Cd$ is the sensitivity of $C$ with respect to a given entry of the correlation matrix.
\item[Step 2f] Set $\Ud_i = \phi(Z_i)\, \Zd_i$, $i=1,\ldots, N$.
\item[Step 3f] Set $\Xd_i = \Ud_i \,/\, m_i(X_i)$,  $i=1,\ldots, N$.
\item[Step 4f] Calculate $\displaystyle \Pd = \sum_{i=1}^N \frac{\partial P}{\partial X_i}\,\Xd_i$~.
\end{itemize}
Here $\phi(x) \equiv {\partial \Phi(x)}/{\partial x}$ is the standard normal probability density function, and
$m_i(x) \equiv \partial M_i(x)/\partial x $ is the probability density function associated with the marginal $M_i(x)$ of the 
$i$-th random factor.

As anticipated, the computational cost of the Forward Pathwise Derivative method scales 
linearly with the number of sensitivities computed $N_\theta$, i.e., the same scaling of finite difference 
approximations of the derivatives $\partial_{\theta_k} E_{\mathbb Q}[P(X)]$. As a result in many situations, 
typically involving complex payouts, the standard implementation of the Pathwise Derivative method 
offers a limited computational advantage with respect to Bumping \cite{aad1}.

In contrast, AAD allows in general a much more efficient implementation of the Pathwise Derivative estimators (\ref{pwd}).  
Indeed, as an immediate consequence of the computational complexity results introduced in the previous Section, 
it can be shown \cite{aad2} that AAD allows the simultaneous calculation of the Pathwise Derivative 
estimators for {\em any} number of sensitivities at a computational cost which is a small multiple (of order 4) of the 
cost of evaluating the original payout estimator. As a result, one can calculate the MC expectation of an arbitrarily 
large number of sensitivities at a {\em small fixed cost}.

\begin{figure}
\begin{center}
\includegraphics[width=80mm,]{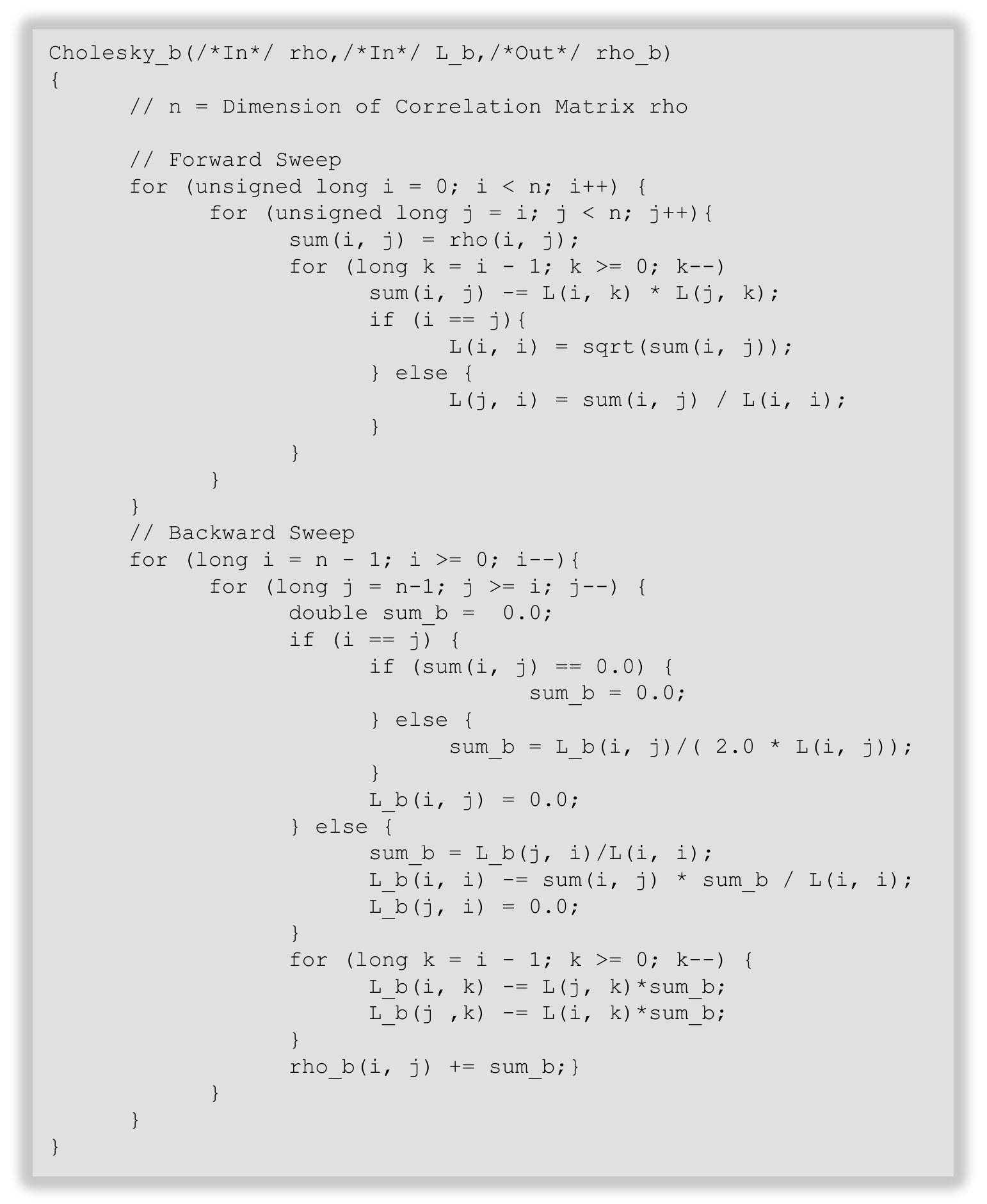}
\end{center}
\vspace{-4mm} \caption{\label{FigCholesky} Adjoint of the Cholesky factorization. The Forward sweep is an exact 
replica of the original factorization.}
\end{figure}

Although AAD can be applied for virtually any model and payout function of interest 
in Computational Finance -- including path-dependent and Bermudan options -- 
here we will concentrate on the  calculation of correlation sensitivities in a Gaussian copula 
framework.  In general, for the reasons mentioned in the previous Section, the AAD implementation of the Pathwise derivative method contains a {\em forward sweep} 
-- reproducing the steps followed in the calculation of the estimator of the option value $P(X)$ -- and a {\em backward sweep}. 
As a result, the adjoint algorithm consists of adjoint counterparts for each of the Steps 1-4 
above executed in reverse order, plus the adjoint of the Cholesky factorization. 

The first step consists in the evaluation of the adjoint of step 4 of the Forward sweep, calculating the 
derivatives of the Payout with respect to the components of the state vector
\begin{equation}
\bar X_k = \frac{\partial P(X)}{\partial X_k}~,
\end{equation}
with $k=1,\ldots,N$. These derivatives can be calculated efficiently using AAD, as discussed in Ref.~\cite{aad1}.

In turn, the adjoint of Step 3 of the Forward sweep is given by 
\begin{equation}
\bar U_k = \bar M^{-1}_k(U_k, \bar X_k) =  \frac{\bar X_k}{m_k(X_k)}~, 
\end{equation}
for $k =1,\ldots, N$.
The vector $\bar U$ is then mapped into the adjoint of the correlated standard normal variables $\bar Z$ through the 
counterpart of Step 2 
\begin{equation}
\bar Z_k = {\bar \Phi(Z_k, \bar U_k)} = {\bar U_k}\,{\phi(Z_k)}~.
\end{equation}
The adjoint of Step 1 performing the matrix vector product $Z = C \tilde Z$ reads
\begin{equation}
\bar C_{i,j} = \sum_{k=1}^N \frac{\partial Z_k}{\partial C_{i,j}}\, \bar Z_k = \tilde Z_j\, \bar Z_i
\end{equation}
or $\bar C = \bar Z \tilde Z^t$.
By applying the chain rule, it is straightforward to realize that the adjoint $\bar w$ of each intermediate variable $w$
in the succession of Steps 0-4 represents the derivative of the Payout estimator with respect to $w$, or $\bar w = \partial P / \partial w$.
In particular the quantities $\bar C_{i,j}$ calculated at the end of the adjoint of Step 1 represent the derivatives of 
the payout estimator with respect to the the entries of the triangular Cholesky matrix
$C$, namely the pathwise estimator (\ref{pwd}) with $\theta_k = C_{i,j}$. 

In summary, the AAD implementation of the Pathwise Derivative Estimator consists of Step 1-4 described above (forward sweep) 
plus the following steps of the backward sweep: 
\begin{itemize} 
\item[Step 5] Evaluate the Payout adjoint $\bar X_k = \partial P/\partial X_k$, for $k=1,\ldots,N$.
\item[Step 6] Calculate $\bar U_k = {\bar X_k}/{m_k(M^{-1}_k(U_k))}$, $k=1,\ldots,N$.
\item[Step 7] Calculate $\bar Z_k = {\bar U_k}{\phi(Z_k)} $, $k=1,\ldots,N$. 
\item[Step 8] Calculate $\bar C = \bar Z \tilde Z^t$.
\end{itemize}

At this point in the calculation, there is an interesting complication.
The natural AAD approach would average the values of $\bar C$ from each of the
MC paths. This average $\bar C$ can be converted into derivatives
with respect to the entries of the correlation matrix $\rho$ by means of the adjoint of the Cholesky factorization \cite{Smith}, namely
a function of the form
\begin{equation}
\bar \rho = \texttt{CHOLESKY}\_\texttt{B}(\rho, \bar C)
\end{equation}
providing
\begin{equation}
\label{rhobar}
\bar \rho_{i,j} = \sum_{l,m=1}^N \frac{\partial C_{l,m}}{\partial \rho_{i,j}}\bar C_{l,m}~.
\end{equation}
The pseudocode for the adjoint Cholesky factorization is given in Fig.~\ref{FigCholesky}.  
By inspecting the structure of the pseudocode it appears clear that its computational cost is just a small multiple (of order 2)
of the cost of evaluating the original factorization. Indeed, the adjoint algorithm essentially contains 
the original Cholesky factorization plus a backward sweep with the same complexity and a similar number of operations. 

The complication with this implementation is that it gives an estimate for the correlation risk, but it does not provide a corresponding
confidence interval.  An alternative approach would be to convert $\bar C$ to $\bar \rho$ for each individual path,
and then compute the average and standard deviation of $\bar \rho$ in the usual way.  However, the numerical results
will show that this is rather costly.  An excellent compromise between these two extremes is to divide the $N_{MC}$ paths
into $N_b$ 'bins' of equal size.  For each bin, an average value of $\bar C$ is computed and converted into 
a corresponding value for $\bar \rho$.  These $N_b$ estimates for $\bar \rho$ can then be combined in the usual way
to form an overall estimate and confidence interval for the correlation risk.

The computational benefits can be understood by considering the computational costs for both the standard evaluation
and the adjoint Pathwise Derivative calculation.  In the standard evaluation, the cost of the Cholesky factorization
is $O(N^3)$, and the cost of the MC sampling is $O(N_{MC} N^2)$, so the total cost is $O(N^3 + N_{MC} N^2)$.
Since $N_{MC}$ is always much greater than $N$, the cost of the Cholesky factorization is usually negligible.
The cost of the adjoint steps in the MC sampling is also $O(N_{MC} N^2)$, and when using $N_b$ bins the
cost of the adjoint Cholesky factorization is $O(N_b N^3)$.  To obtain an accurate confidence interval, but with
the cost of the Cholesky factorisation being negligible, requires that $N_b$ is chosen so that
$1 \ll N_b \ll N_{MC} / N$.  Without binning, i.e., using $N_b = N_{MC}$, the cost to calculate the average 
of the estimators (\ref{rhobar}) is $O(N_{MC} N^3)$, and so the relative
cost compared to the evaluation of the option value is $O(N)$.

\begin{figure}
\begin{center}
\hspace{-2mm}
\includegraphics[width=80mm,angle=0]{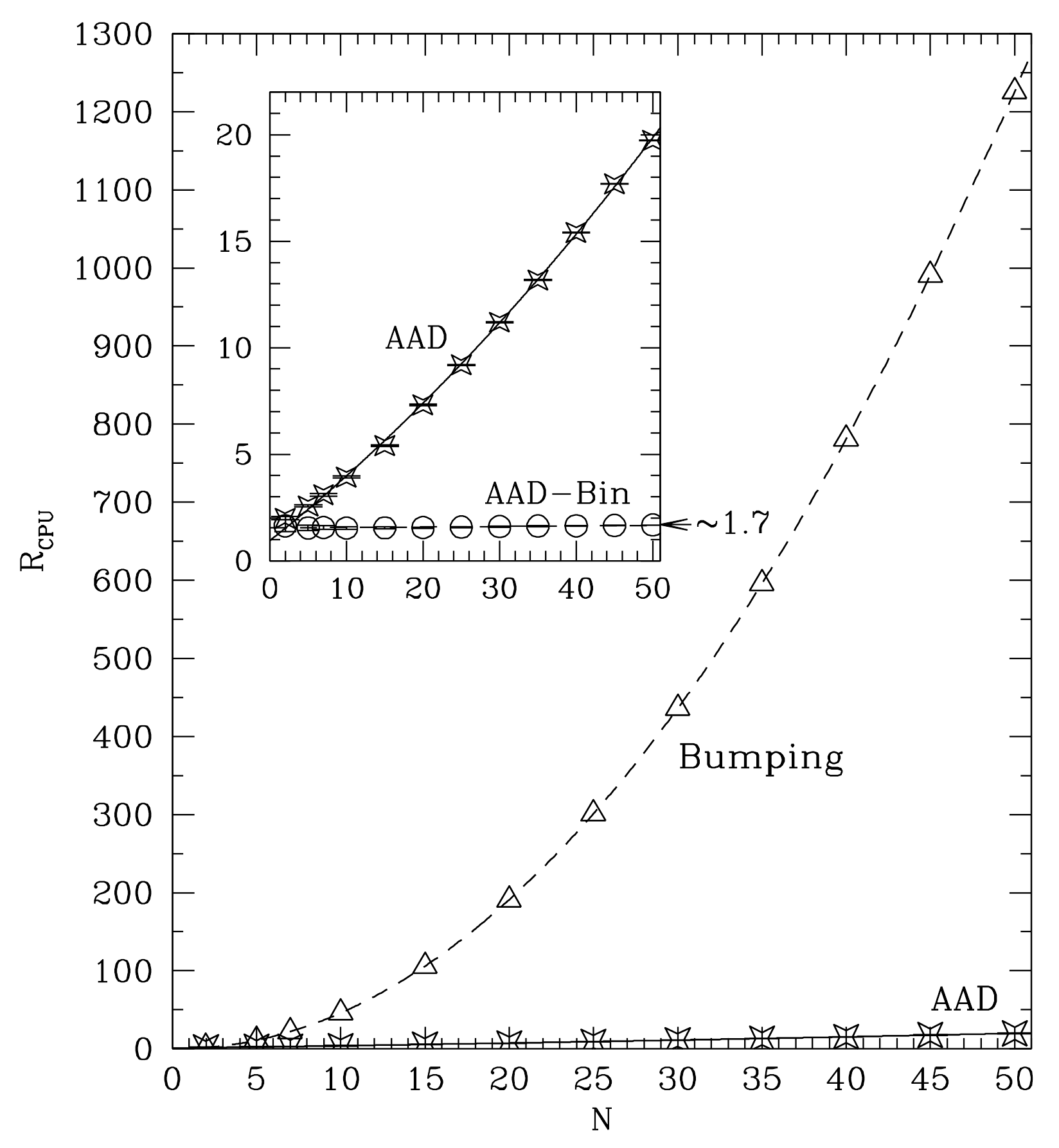}
\end{center}
\vspace{-5mm} \caption{\label{cpuratio} Ratios of the CPU time required for the calculation of the
option value, and correlation Greeks, and the CPU time spent for the computation of the value alone, as functions of the
number of names in the basket, for $N_{MC} = 10^5$. Symbols: Bumping (one-sided finite differences) (triangles), AAD 
without binning (i.e.~$N_b=N_{MC}$) (stars), AAD with binning ($N_b=20$)
(empty circles).  Lines are guides for the eye, and the MC uncertainties are smaller than the symbol sizes.  }
\end{figure}

The binning procedure described above can be generalized to any situation in which the standard 
solution procedure involves a common preprocessing step before any of the path calculations are 
performed.  Other examples would include calibration of model parameters to market prices, 
or a cubic spline construction of a local volatility surface.  In each case, there is a linear
relationship between the forward mode sensitivities before and after the preprocessing step, and
therefore a linear relationship between the corresponding adjoint sensitivities.

The algorithm described above can be applied whenever the option pricing problem can
be formulated as an expectation value over a set of random factors whose distribution 
is modelled as a Gaussian copula. This include in general a variety of Basket Options 
common across all asset classes, or structured swaps whose coupon depends on a specific observation
of a set of correlated rates. In addition, the same ideas can be extended to the simulation of correlated diffusion
processes \cite{aad2}.

\section*{Numerical Tests}

As a numerical test ground we consider the case of 
Basket Default Options \cite{ChenGlass}. In this context, the random factors $X_i$ represent the default 
time $\tau_i$ of the $i$-th name, e.g., the time a specific company in a reference pool of $N$ names 
fails to pay one of its liabilities as specified by the terms of the contract priced. 
In particular, in a $n$-th to default Basket Default Swap
one party (protection buyer) makes regular payments to a counterparty (protection seller) 
at time $T_1,\ldots,T_M \leq T$ provided that less than $n$ defaults events among the components of the basket are observed before time $T_M$. 
On the other hand, if $n$ defaults occur before time $T$, the regular payments cease and the 
protection seller makes a payment to the buyer of $(1-R_i)$ per unit notional, where $R_i$ is the 
normalized recovery rate of the $i$-th asset.
The value at time zero of the Basket Default Swap on a given realization of the default 
times $\tau_1,\ldots,\tau_N$, i.e., the Payout function, can be therefore expressed as
\begin{equation}
P(\tau_1,\ldots,\tau_N) = P_{prot}(\tau_1,\ldots,\tau_N)-P_{prem}(\tau_1,\ldots,\tau_N)  
\end{equation}  
i.e., as the difference between the so-called protection and premium legs.
The value leg is given by
\begin{equation}\label{prem}
P_{prot}(\tau_1,\ldots,\tau_N) = (1 - R_n) D(\tau) \mathbb{I}(\tau \leq T)~,
\end{equation}
where $R_n$ and $\tau$ are the recovery rate and default time of the $n$-th to default, respectively, 
$D(t)$ is the discount factor for the interval $[0,t]$ (here we assume for simplicity 
uncorrelated default times and interest rates), and ${\mathbb I}(\tau \leq T)$ is 
the indicator function of the event that the $n$-th default occurs before $T$. 
The premium leg reads instead, neglecting for simplicity any accrued payment,
\begin{equation}\label{prot}
P_{prem}(\tau_1,\ldots,\tau_N) = \sum_{k=1}^{L(\tau)} s_k D(T_k)
\end{equation}
where $L(\tau) = \max [k \in \{1,\ldots, M\} / T_k < \tau ] $, and $s_k$ is the premium 
payment (per unit notional) at time $T_k$.

In order to apply the Pathwise Derivative method to the payout above, the indicator functions in (\ref{prot}) and (\ref{prem}), 
need to be regularized \cite{GlassMCbook,ChenGlass}. One simple and practical way of doing that is to replace the indicator functions  with their 
smoothed counterpart, at the price of introducing a small amount of bias in the Greek estimators. For the problem at hand,
as it is also  generally the case, such bias can be easily reduced to be smaller than the statistical errors that can be obtained 
for any realistic number of MC iteration $N_{MC}$ (for a more complete discussion of the topic of payout regularization see Refs.~\cite{aad1,aad2,Gilesmcqmc}).

The remarkable computational efficiency of AAD is illustrated in Fig.~\ref{cpuratio} for the
Second to Default Swap. Here we plot the ratio of the CPU time  required for the calculation of the
value of the option, and all its pairwise correlation sensitivities, and the CPU time spent for the computation 
of the value alone, as functions of the number of names in the basket. As expected,  for standard finite-difference 
estimators, such ratio increases quadratically with the number of names in the basket. 
Already for medium sized basket ($N\simeq 20$) the cost associated with Bumping is over 
100 times more expensive than the one of AAD.

Nevertheless, at a closer look (see the inset of Fig.~\ref{cpuratio}), the relative cost of AAD without binning
is $O(N)$, for the reasons explained earlier.
However, when using $N_b=20$ bins the cost of the adjoint Cholesky computation is negligible and the
numerical results show that all the Correlation Greeks can be obtained with a mere 70\% overhead compared to 
the calculation of the value of the option. This results in over 2 orders of magnitude savings in computational 
time for a basket of over 40 Names.

\newpage

\section*{Conclusions}

In conclusion, we have shown how Adjoint Algorithmic Differentiation allows an extremely efficient calculation of correlation Risk in Monte Carlo.  
The proposed method relies on using the Adjoint mode of Algorithmic Differentiation to organize the calculation of the Pathwise Derivative estimator, and to
implement the adjoint counterpart of the Cholesky factorization. For any number of underlying assets or names in a portfolio, the proposed method allows the calculation of
the complete pairwise correlation Risk at a computational cost which is at most 4 times the cost of calculating the option value itself, resulting in remarkable computational savings
with respect to Bumping. 
We illustrated the method for a Gaussian copula-based Monte Carlo computation, and we tested it numerically
for Portfolio Default Options.  In this application, the proposed method is 100 times faster than Bumping for 20 names, and over 1000 times for 40 names.  The method generalizes
immediately to other kind of Elliptic copulas, and to a general diffusive setting. In fact, it is a specific instance of a general AAD approach to the implementation of the
Pathwise Derivative method that will be discussed in a forthcoming publication \cite{aad2}.

{\bf Acknowledgments:}
It is a pleasure to acknowledge useful discussions with Alex Prideaux, Adam and Matthew Peacock, 
Jacky Lee and David Shorthouse.  Valuable help provided by Mark Bowles and Anca Vacarescu in the 
initial stages of this
project is also gratefully acknowledged. 
The opinions and views expressed in this paper are 
uniquely those of the authors, and do not necessarily represent those of Credit Suisse Group.

\bibliography{biblio}     

\end{document}